\documentclass[epj]{svjour}
\usepackage{graphics}
\begin{document}
\title{Sub- and above barrier fusion of loosely bound  $^6$Li with $^{28}$Si}
\author{Mandira Sinha\inst{1,2}, H. Majumdar\inst{1,}\thanks{\emph{Email address:} harashit.majumdar@saha.ac.in}, P. Basu\inst{1}, Subinit Roy\inst{1}, R. Bhattacharya\inst{2}, M. Biswas\inst{1}, M. K. Pradhan\inst{1}, R. Palit\inst{3}, I. Mazumdar\inst{3} and S. Kailas\inst{4}
}                     
\offprints{}          
\institute{Saha Institute of Nuclear Physics, 1/AF, Bidhan Nagar, Kolkata-700064, India, \and Gurudas College, Narikeldanga, Kolkata-700054, India \and Tata Institute of Fundamental Research, Homi Bhabha Road, Mumbai-400005, India \and Nuclear Physics Division, Bhabha Atomic Research Centre, Mumbai-400085, India }
\date{Received: date / Revised version: date}
%
\abstract{ Fusion excitation functions are measured for the system $^6$Li+$^{28}$Si using the characteristic $\gamma$-ray method, encompassing both the sub-barrier and above barrier regions, viz., $E_{lab}$= 7-24 MeV. Two separate experiments were performed, one for the above barrier region ($E_{lab}$= 11-24 MeV) and another for the below barrier region ($E_{lab}$= 7-10 MeV). The results were compared with our previously measured fusion cross section for the $^7$Li+$^{28}$Si system. We observed enhancement of fusion cross section at sub-barrier regions for both $^6$Li and $^7$Li, but yield was substantially larger for $^6$Li. However, for well above barrier regions, similar type of suppression was identified for both the systems. 
} 
\maketitle
\section{Introduction}
Theoretical and experimental studies of fusion excitation functions induced by stable loosely bound nuclei like $^6$Li, $^7$Li, $^9$Be at near barrier energies, provide new insights into reaction dynamics and structure effects caused by interplay between fusion, loose structure, transfer and breakup to the continuum \cite{Huss93,Dass96,Hagi00,Diaz,Yosi96,Taka97,Mdas1,Sig9,Trot00,Cbeck,Cbeck1,YW,Navin,Rabe04,Taki93,NZag,Rehm,mukh,man07,man08}. Two of the most interesting and important results are fusion suppression/enhancement \cite{Huss93,Dass96,Hagi00,Diaz,Yosi96,Taka97,Mdas1,Sig9,Trot00,Cbeck,Cbeck1,YW} and weakening of the usual threshold anomaly near the barrier. Fusion augmentation or decrease with respect to the one dimensional barrier penetration model (1D BPM) may be intuitively understood as follows: Due to coupling between relative motion and internal degrees of freedom, like inelastic excitation, nucleon transfer, breakup, collective vibration, static deformation etc., the single barrier evolves into a multiple barrier distribution. Consequently enhancement of fusion will occur owing to lowering of the effective barrier. However, in the case of weakly bound (cluster type) projectiles, that have smaller breakup threshold, there is an appreciable probability of breakup leading to lowering of fusion probability. So the competition between these two opposite tendencies ultimately decides the quantitative nature of the fusion excitation function.

Though there exist a number of precise measurements of fusion of stable weakly bound projectiles with heavy mass targets, there are only few fusion data with medium mass targets (A$\sim$ 60) at near barrier energies \cite{Cbeck,SMores,Gomes04,Adiaz5,Ipadro}. But almost all of these address to above barrier fusion. In this region mention may be made of fusion measurement at near barrier energies for $^{6,7}$Li+$^{59}$Co by C. Beck $et$ $al$. \cite{Cbeck} showing good agreement with 1D BPM prediction at energies above the barrier. However close to the Coulomb barrier this work observed slight enhancement of $^{6}$Li induced fusion compared to that with $^{7}$Li.

In the relevant scenario of target mass range A$\sim$ 20-30 there are very few experimental attempts \cite{man08,Mray,Pakou09} at sub-barrier energies and few data exist at energies above the Coulomb barrier \cite{man07,Ipadro,Mray,Pakou09,Hugi,SKal90,GMar05,Kali06}. However the present authors made a consistent attempt \cite{man07,man08} to experimentally explore the fusion behaviour for $^7$Li+$^{28}$Si starting from the sub barrier region to well above (up to $\sim$ 3$V_b$) the barrier. We mainly observed the signature of fusion enhancement below the barrier and small hindrance at energies twice that of the barrier, findings which are similar to the theoretical predictions \cite{Dass96,Hagi00} for the systems $^{11}$Li+$^{208}$Pb and $^{11}$Be+$^{208}$Pb respectively. It is interesting to note that most of the experiments \cite{Pakou09,Hugi,SKal90,GMar05,Kali06} conducted in this mass region observed no effect of coupling on fusion yields.

 As the attempts at measuring sub-barrier fusion are very meagre, more experiments are necessary to understand the reaction dynamics of loosely bound projectiles not only at sub barrier energies but also at higher energies. Moreover $^6$Li being more loosely bound than $^7$Li it will be instructive and worthwhile to compare coupling effects on fusion in these two cases. With this objective in view, we present here the experimental measurement of fusion of $^{6}$Li with $^{28}$Si at sub-barrier and above barrier energies, $E_{lab}$= 7-24 MeV, using the characteristic $\gamma$-ray method. 
 
\section{Experimental Details}

The range of bombarding energy was covered by doing two separate experiments. First we measured the total fusion (TF) cross sections for $^6$Li+$^{28}$Si at above barrier energies viz., $E_{lab}$= 11, 12, 14, 16, 18, 20, 22 and 24 MeV. This experiment was conducted at the 14 MV Pelletron accelerator facility of TIFR-BARC (at Mumbai) with $^6$Li (3$^+$) beams and current varying between 2-10 pnA. A small thin walled aluminium chamber was used to house the target. The target consisted of 192 $\mu$g/cm$^2$ natural silicon sandwiched between two thin gold layers (40 $\mu$g/cm$^2$ and 100 $\mu$g/cm$^2$) in order to prevent oxidation and was prepared using vacuum evaporation technique. The average energy loss in the target is about 200 keV. The characteristic $\gamma$-rays emitted from the evaporation residues were detected using a Compton suppressed Clover detector placed at 55$^o$ with respect to the beam direction. The detector resolution achieved was better than 2.8 keV for the 1408 keV line of $^{152}$Eu. The details of experimental setup and efficiency of the detector was discussed in our previous measurement \cite{man07}. 

For sub and near barrier energy region, we performed another experiment at the 3 MV Pelletron accelerator centre (IOP, Bhubaneswar) with $^6$Li (2$^+$, 3$^+$) beam (3-25 pnA) at energies, $E_{lab}$= 7, 8, 9 and 10 MeV. A self supported thin target of $^{28}$Si (175 $\mu$g/cm$^2$) and a specially designed small thin walled target chamber made of stainless steel was used to measure the fusion cross section using the characteristic $\gamma$-ray method. It may be mentioned that target thickness was measured with the $\alpha$ energy loss method with a three-line $\alpha$ source and uncertainty including the effects of contaminants, was estimated to be about 5$\%$. Here also the $\gamma$-rays emitted from the evaporation residues were detected using a HPGe detector placed at 125$^o$ with respect to the beam direction. The resolution of the detector was found to be 2 keV for the 1408 keV line of $^{152}$Eu. The experimental setup and efficiency of the detector are explained in our earlier work with $^7$Li+$^{28}$Si \cite{man08}.

\section{Analysis}

The evaporation residues resulting from the fusion of $^6$Li +$^{28}$Si were identified by their characteristic $\gamma$-rays. Gamma cross sections (${\sigma}_{\gamma}$) were extracted after analysing the $\gamma$-ray spectra and using relevant efficiencies, beam and target specifications etc., through the formula  $\sigma_{\gamma}$= $N_{\gamma}$/[$\epsilon_{\gamma}$. $N_B$. $N_T$] where $N_{\gamma}$= No. of counts in the ${\gamma}$-peak, $\epsilon_{\gamma}$= full energy detection efficiency of ${\gamma}$-ray. $N_B$ and $N_T$ are respectively total no. of incident particles and no. of target nuclei per unit area. This formulation is independent of origin of residue formation (CF, ICF or transfer). It may be mentioned that these experimental ${\gamma}$- cross sections take into account all ${\gamma}$-transitions from all residues originating from possible sources like CF, ICF or residues formed after transfer. The important residues at above barrier energies ($E_{lab}$=11-24 MeV) are $^{29}$Si, $^{32}$S, $^{28}$Si, $^{31}$P and $^{26}$Al. The cumulative contributions from all of the above channels account for nearly 83-90$\%$ of the total fusion cross section, as estimated from the statistical model code CASCADE \cite{six}. Main contribution to fusion come from $\alpha$$p$+$^{29}$Si, $p$$n$+$^{32}$S, $d$+$^{31}$P and $\alpha$$d$+$^{28}$Si channels. Some of the prominent identified $\gamma$-rays are 1.273+1.266 MeV ($^{29}$Si+$^{31}$P), 2.028 MeV ($^{29}$Si), 2.230+2.233 MeV ($^{32}$S+$^{31}$P), 1.779+1.794 MeV ($^{28}$Si+$^{29}$Si), 0.416 MeV ($^{26}$Al). As natural silicon has contaminants $^{29}$Si (4.68 $\%$) and $^{30}$Si (3.08 $\%$), their contributions were estimated using CASCADE and this yielded an over all maximum error of about 5 $\%$ in the total fusion cross section. This is taken into account in estimation of total uncertainty in fusion cross section.

\begin{figure}
\resizebox{0.5\textwidth}{!}
{\includegraphics{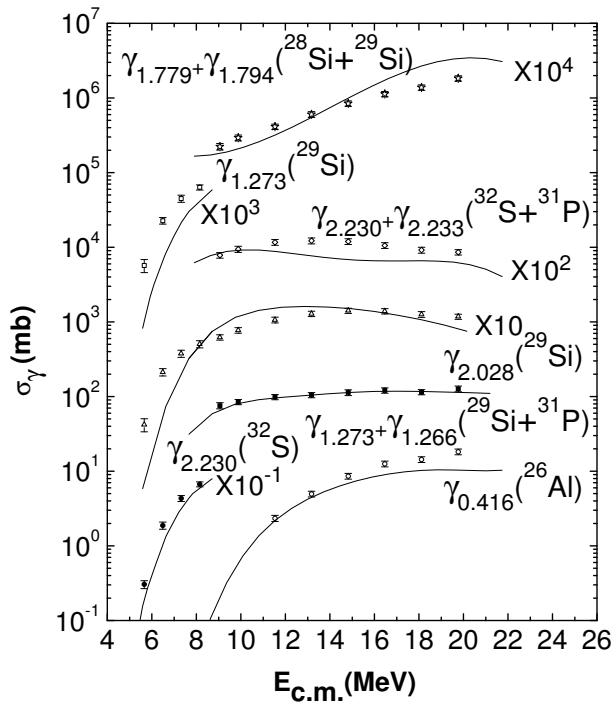}}
\caption{ Measured $\gamma$-ray cross sections, calculated from the yield of the $\gamma$-ray from de-exciting residues, plotted against $E_{c.m.}$. The solid lines show the fit obtained from the statistical model code CASCADE.}
\label{6lisi-tifr}      
\end{figure}

The measured $\gamma$-ray cross sections at different projectile energies in $E_{c.m.}$ are compared with the CASCADE estimation and are shown in Fig.~\ref{6lisi-tifr}. It appears that reasonable fits with the statistical model predictions are obtained for most of the $\gamma$-ray cross sections signifying that most these $\gamma$-transitions are from residues of complete fusion component. The exceptions were found, e.g., for the $\gamma$-rays 1.779+1.794 MeV at the higher energy region where over prediction of the data has occurred and also for the observed $\gamma$-rays of energies 2.028 MeV and 1.273 MeV showing definite under predictions at the near and sub-barrier energies.

\begin{figure}
\resizebox{0.5\textwidth}{!}
{\includegraphics{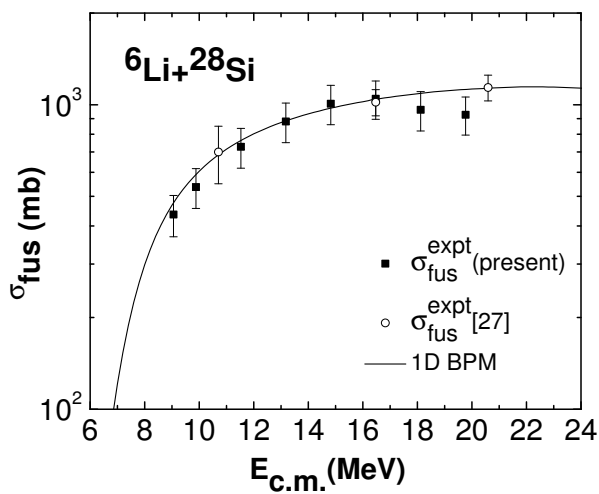}}
\caption{ Fusion excitation function measured at above barrier energies
($E_{c.m.}$= 9.06-19.76 MeV) for the present system $^6$Li+$^{28}$Si. The data points for each energy are shown as solid squares. The solid curve shows the fit obtained with 1D BPM calculations using the code CCFULL in the no coupling mode. Experimental fusion data for the same system, measured by \cite{Hugi}, are also shown (open circles) for comparison.}
\label{Fus-above}      
\end{figure}

The total $\gamma$-ray cross section was obtained by summing over all the above mentioned $\gamma$-ray cross sections. The total fusion cross section was then estimated as the ratio of the above total $\gamma$-ray cross sections and the total branching factor $F_\gamma$ which was estimated using the code CASCADE following the standard and well established method widely cited in literature, e.g., Refs.~\cite{mukh,Sch}. As $\gamma$ cross sections are measured using transitions from all possible residues originating from possible sources like CF, ICF or residues formed after transfer, fusion cross section also contains the contribution from possible ICF/transfer component. The calculated value of $F_\gamma$ varies from 46$\%$ to 57$\%$ in the energy region under study. The uncertainty of $F_\gamma$ was estimated to be of the order of 10$\%$. With due consideration for statistical error of $\gamma$-ray yield, absolute efficiency of the detectors, systematic error in target thickness measurement and integrated beam current \cite{man07}, we obtained the total fusion cross section uncertainty to be varying between 14$\%$ to 15.5$\%$. The total fusion excitation at above barrier energies is shown in Fig. \ref{Fus-above}. We compared our fusion data with 1D BPM calculations obtained from the code CCFULL \cite{Ccfull} used in the no coupling mode, where the optical model parameters were taken from \cite{man07}. The CCFULL calculation on the present system $^6$Li+$^{28}$Si, yielded a value for the barrier of $V_b$= 6.87 MeV. The present fusion data was also compared with the existing fusion measurement of the same system by Hugi $et$ $al$. \cite{Hugi}, at energies above the barrier.

Sub and near barrier data, experimentally found for the first time, were analysed and  presented separately, as here we have non-overlapping characteristic $\gamma$-rays permitting individual channel cross section measurement and also extraction of fusion excitation from single $\gamma$-ray cross section. In the energy regime under investigation ($E_{lab}$= 7-10 MeV), the most important residues detected are $^{29}$Si and $^{32}$S. The contributions of these observed channels $^{29}$Si and $^{32}$S are almost about 78-80$\%$ of the total fusion cross section for $^6$Li+$^{28}$Si system, as estimated from CASCADE, in the same energy region. The prominent identified $\gamma$-rays are 1.273 MeV ($^{29}$Si), 2.028 MeV ($^{29}$Si) and 2.230 MeV ($^{32}$S). The estimated projectile energy loss in the half thickness target is about 123 to 100 keV in the energy regime 7-10 MeV. These factors were taken into account and $\gamma$-ray cross sections were plotted as a function of effective projectile energy and shown along with CASCADE predictions in Fig.~\ref{6lisi-tifr}. 
 
The total fusion cross section was extracted as the ratio of the total experimentally measured $\gamma$-ray cross sections and the corresponding total branching factor $F_{\gamma}$ evaluated from the code CASCADE. The calculated value of $F_\gamma$ varies from 42$\%$ to 50$\%$ in the energy region under review. The measured fusion cross sections at sub and near barrier region are plotted in Fig.~\ref{fustot-1273} and the over all resulting uncertainty in projectile energy is also shown. The uncertainty in the measurement of the fusion cross section was estimated to be about 15-18$\%$ for all energies, except for the lowest energy, where it was nearly 25$\%$, due to very poor statistics.  
  
We also measured the total fusion cross section by a second method i.e., by summing the measured channel cross sections of the observed residues. Each channel cross section was extracted by dividing the measured $\gamma$-ray cross sections of the individual channel by their corresponding branching factor. The fusion excitation function from these channel 
cross sections are evaluated and also plotted in Fig.~\ref{fustot-1273}. As fusion cross section can be extracted from measured $\gamma$-ray cross section of a single residue provided the channel is intensely populated we found the fusion excitation from the measured cross section of $\gamma_{2.230}$ of residue $^{32}$S and the corresponding branching factor. These results are also plotted in Fig.~\ref{fustot-1273}. For more clarity and quantitative characterization we present the values of few cross sections of prominent $\gamma$- rays ($\sigma_{\gamma}$), residue cross sections ($\sigma_{ch}$) and total fusion cross sections ($\sigma_{TF}$) measured at sub- and near barrier energies in Table \ref{table1}.

\section{Discussion}

The measured fusion cross sections at above barrier energies (Fig.~\ref{Fus-above}), showed good agreement with 1D BPM estimates and the previous measurement by \cite{Hugi} using the evaporation $\alpha$ measurement method, except at two higher energies points at $E_{c.m.}$$\geq$ 2$V_b$. In this higher energy region 1D BPM overpredicts our experimental 
fusion data  by about 12-17 $\%$, almost similar to our previous observation in the case of fusion of $^7$Li with $^{28}$Si \cite{man07}. Similar observations are also reported by Kovar $et$ $al$. \cite{kover} and Takahashi $et$ $al$. \cite{Taka97}. Fusion behaviour with loosely bound systems has not been earlier explored in this high energy region. However calculations for heavier systems \cite{Diaz}, experimental measurements of Signorini $et$ $al$. \cite{Sig9} and Dasgupta $et$ $al$. \cite{Mdas1} were done up to at most E= 1.7 $V_b$ only. They reported gradual saturation of ICF with increasing energy. Our results are also similar up to 2$V_b$, but beyond this energy we observe some sort of decrease in fusion cross section (compared to 1D BPM). As our earlier work with $^7$Li+$^{28}$Si \cite{man07} yielded almost similar values for  fusion cross section at higher energy both with $\gamma$ method and $\alpha$ method (mostly measuring CF fusion), we conjecture that here in this case also fusion decrease might be due to smaller probability of ICF formation with break up component. Also in this region fusion excitation changes very slowly with bombarding energy and this suggests that interaction barrier has less effect on fusion phenomenon relevant here.

\begin{figure}
\resizebox{0.5\textwidth}{!}
{\includegraphics{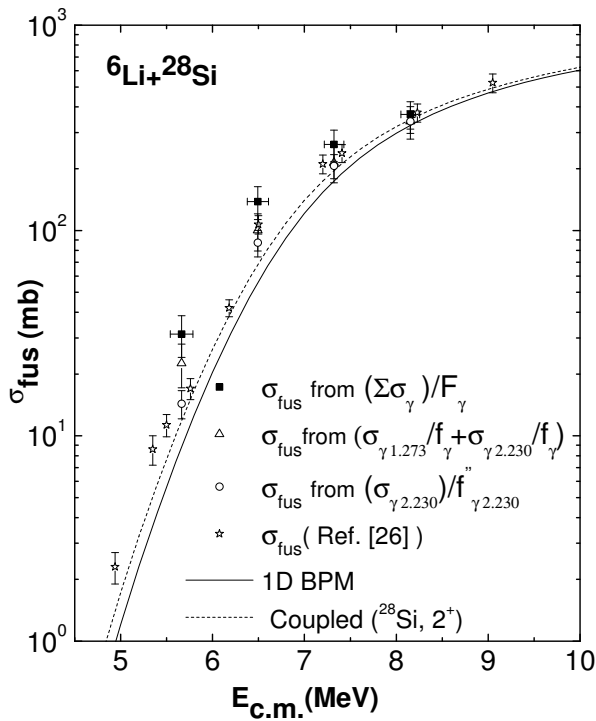}}
\caption{ Same as Fig.~\ref{Fus-above}, but for sub and near barrier energy region ($E_{c.m.}$= 5.66-8.15 MeV). Solid squares represent the fusion cross section obtained from summing of all the experimentally measured $\gamma$-cross sections ($\sigma_{\gamma}$). Open triangles and circles represent cross section values obtained from summing of channel cross sections ($\sigma_{ch}$) and from the cross section of 2.230 MeV $\gamma$-ray of $^{32}$S residue; open star represent the estimated fusion cross section by \cite{Pakou09}. See text for explanation. Solid curve gives the fit obtained with 1D BPM calculations using the code CCFULL in the no coupling mode while the dashed line gives the estimation considering coupling to the 2$^+$ (1.779 MeV) state of the target.}
\label{fustot-1273}      
\end{figure}

\begin{table*}
\caption{\label{table1} Measured cross sections of prominant ${\gamma}$-rays ($\sigma_{\gamma}$), important residues ( $\sigma_{ch}$) and total fusion ($\sigma_{TF}$) of $^6$Li+$^{28}$Si at near barrier energies.}
\vglue 0.4cm
\begin{center}
\begin{tabular}{|c|c|c|c|c|c|}
\hline 
& & & & & \\
$E_{c.m.}$&$\gamma$-energy &$\sigma_{\gamma}$ &Channel  &$\sigma_{ch}$  &$\sigma_{TF}= \frac{\sum{\sigma_{\gamma}}}{F_{\gamma}}$\\
(MeV)&(MeV) &(mb)& &(mb) & (mb)\\
\hline
& & & & & \\
5.66& 2.230 &3.05 $\pm$ 0.4 & $^{32}$S &6.38 $\pm$ 1.0& 31.3 $\pm$ 6.2\\
& 1.273 &5.7 $\pm$ 1.1  & $^{29}$Si &16.17 $\pm$ 3.2 &\\
& 2.028 & 4.2 $\pm$ 0.8 & $^{29}$Si &17.8 $\pm$ 4.0 &  \\
& & & & & \\
6.49& 2.230 &18.77 $\pm$ 2.0& $^{32}$S &36.80 $\pm$5.4& 138.4 $\pm$ 25\\
& 1.273 & 22.73 $\pm$ 2.3 & $^{29}$Si &63.24 $\pm$ 9.0 &\\
& 2.028 & 21.26 $\pm$ 2.3& $^{29}$Si &73.97 $\pm$ 11 &\\
& & & & & \\
7.32& 2.230 & 43.2 $\pm$ 3.9 & $^{32}$S &81.55 $\pm$ 11& 262.3 $\pm$ 46\\
& 1.273 & 44.8$\pm$ 4.9 & $^{29}$Si &132.02 $\pm$ 20 &\\
& 2.028 & 37.54$\pm$ 4.1 & $^{29}$Si &112.53 $\pm$ 17 & \\
& & & & & \\
8.15& 2.230 & 66.9 $\pm$ 5.2& $^{32}$S &123.9 $\pm$ 15& 367.8 $\pm$ 56\\
& 1.273 & 63.6 $\pm$ 5.1& $^{29}$Si &216.11 $\pm$ 28 &\\
& 2.028 & 50.4 $\pm$ 5.5 & $^{29}$Si &136.32 $\pm$ 20&\\
\hline
\end{tabular}
\end{center}
\end{table*}

The experimental fusion data at sub- and near barrier region (5.5 MeV$\leq$$E_{c.m.}$$\leq$8.5 MeV) obtained by the two methods as discussed in the earlier section, are remarkably consistent within uncertainties, as shown in Fig.~\ref{fustot-1273}. The fusion data are compared with the 1D BPM calculations and also with the theoretical estimation using the coupled channel code CCFULL where rotational coupling with 2$^+$ state (1.779 MeV) of the target was taken into account. The 1D BPM estimate predicts the data only at two energy points above $E_{c.m.}$$>$ 6.8 MeV, but it grossly underpredicts the sub-barrier region. These results are similar to sub-barrier fusion behaviour obtained by M. Ray $et$ $al$. \cite{Mray}. Enhanced fusion yield at these energies, as compared to 1D BPM calculations, varies between 72-34$\%$ in the region $E_{c.m.}$= 5.66-8.15 MeV. Introduction of rotational coupling does show some enhancement but can not fully explain the observed enhancement. Very recently Pakou $et$ $al$. \cite{Pakou09}, have estimated the fusion cross section of $^6$Li+$^{28}$Si by subtracting the theoretically fitted transfer cross sections (assumed to be total direct component) from the measured total reaction cross sections. For comparison these findings are also shown in Fig.~\ref{fustot-1273}. It is seen that our fusion measurements with a single residue $^{32}$S agree well with the estimates of Ref.~\cite{Pakou09}. In the present measurement we observed that $\alpha$+$p$+$^{29}$Si* channel is strongly populated in the near barrier and sub-barrier energy region. This has also been observed in exclusive $\alpha$-$\gamma$ and $p$-$\gamma$ coincidence measurements by Pakou $et$ $al$. \cite{Pak1}. This exit channel $\alpha$+$p$+$^{29}$Si* can be populated primarily by (i) complete fusion of $^6$Li+$^{28}$Si, (ii) incomplete fusion of the deuteron after breakup of $^6$Li or after the deuteron transfer to particle unbound state i.e., $^6$Li+$^{28}$Si $\rightarrow$ $\alpha$+$^{30}$P* $\rightarrow$ $\alpha$+$p$+$^{29}$Si*, (iii) $^6$Li+$^{28}$Si $\rightarrow$ ($^5$Li)+$^{29}$Si* $\rightarrow$ $\alpha$+$p$+$^{29}$Si*, the 1-$n$ transfer to $^{28}$Si and subsequent breakup of unbound $^5$Li to $\alpha$ and proton. The arguement put forward by Pakou $et$ $al$. \cite{Pak1} to exclude the process (ii) in explaining the $\alpha$-$\gamma$ coincidence data, is not very conclusive. We feel that $\alpha$-particle yield in coincidence with the characteristic $\gamma$-rays of $^{29}$Si will also include the contribution of process (ii). The statistical model calculations with the codes CASCADE and PACE2 have shown that for $d$-ICF (assuming deuteron is moving with beam velocity) process, $^{29}$Si+$p$ is the most dominating channel. The residue $^{29}$Si accounts for 92 to 95$\%$ of the $d$-incomplete fusion process in the energy range 5.5 MeV$\leq$ $E_{c.m.}$ $\leq$ 8.5 MeV. In this context, it is to be mentioned that both CASCADE and PACE2 statistical model calculations do not indicate the population of $^{26}$Al and $^{25}$Mg residue channels in this energy range. In our measurement we also have not observed the $\gamma$-rays of $^{26}$Al and $^{25}$Mg. The $d$-incomplete fusion seems to be a favoured channel at low energies as it populates the intermediate nucleus $^{30}$P ($d$+$^{28}$Si$\rightarrow$$^{30}$P$^*$; Q=11.85 MeV) with higher excitation energy compared to $d$-transfer which has an optimum Q-value of -1.65 MeV for the reaction. This has been demonstrated by F.A. Souza $et$ $al$. \cite{Souza} in the two-body kinematics description of the behaviour of excitation energy of the intermediate nuclei, formed in the collision of $^6$Li and $^{59}$Co, as a function of detection angle. The authors have shown that the excitation energies of the intermediate nuclei produced from both $d$- and $\alpha$-incomplete fusion follow the same behaviour as functions of detection angles highlighting the existence of incomplete fusion process at low energies. They have also observed that $d$-incomplete fusion, compared to $\alpha$-incomplete fusion, occurs with higher probability as the former faces lower Coulomb barrier than the latter. This results in higher $\alpha$ yield in the exit channel an observation that corroborates with the conclusion of Ref. \cite{Asriv}. This is also the case for the present study as $^{29}$Si from $d$-incomplete fusion has much higher yield than $^{32}$S from $\alpha$-incomplete fusion at the same bombarding energy. The competition between the two processes i.e., $d$-ICF/$d$-transfer and 1-$n$ transfer, can change with decreasing target mass, thus it is not altogether justified to rule out the contribution of any one of the two processes. However the characteristic $\gamma$-ray detection technique is not capable of identifying the two processes. Hence, we feel that the large population of $^{29}$Si at low energies may have contributions from all the three processes.

\begin{figure}
\resizebox{0.5\textwidth}{!}
{\includegraphics{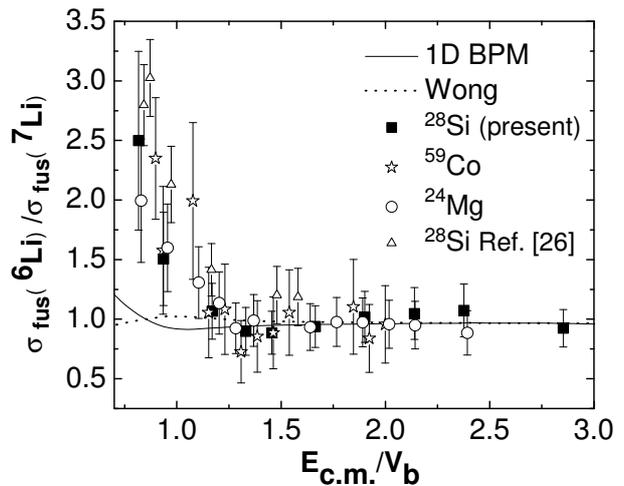}}
\caption{Ratio of measured fusion cross sections for $^6$Li and $^7$Li \cite{man07,man08} projectiles with $^{28}$Si target against $E_{c.m.}$/$V_{b}$ are represented by solid squares. Solid line gives the 1D BPM prediction while the dotted line shows results obtained from Wong's prescription. The same ratios for $^{59}$Co \cite{Cbeck,Adiaz5}, $^{24}$Mg \cite{Mray} and $^{28}$Si \cite{Pakou09} are also shown by open stars, open circles and open traingles respectively.}
\label{7Li+6Li-coup}      
\end{figure}

We have compared the present fusion data of $^6$Li+$^{28}$Si, with our earlier measurement of fusion for $^7$Li+$^{28}$Si system \cite{man07,man08} by the same detection technique, as a function of $E_{c.m.}$/$V_b$, in Fig.~\ref{6,7Li-new1DBPM}. The 1D BPM estimation for both the systems are also shown in Fig.~\ref{6,7Li-new1DBPM}, where solid and dashed curves corresponds to $^7$Li+$^{28}$Si and $^6$Li+$^{28}$Si system, respectively. We observed 
larger fusion yield of the present system $^6$Li+$^{28}$Si, compared to $^7$Li+$^{28}$Si, at close to and below the barrier in the region $E_{c.m.}$/$V_{b}$= 0.82-1.17. However above the barrier region upto 2$V_b$, we have found good agreement of our measured fusion data for both the systems, with the 1D BPM predictions. A small suppression, of about 
similar amount, was apparent beyond the 2$V_b$ region for both the systems. Comparative study by plotting the ratio [{$\sigma_{fus}$}({$^6$Li})/{$\sigma_{fus}$}({$^7$Li})] of measured total fusion cross section of $^6$Li+$^{28}$Si and $^7$Li+$^{28}$Si as a function of $E_{c.m.}$/$V_{b}$ is shown in  Fig.~\ref{7Li+6Li-coup}. In our measurement, the average value of $E_{c.m.}$/$V_{b}$ for the two different systems are taken where the difference of the above values are within 0.01-0.04 from the lowest to highest energies. The estimated error of the ratio of the cross sections is found to vary between 30-17$\%$ in the energy range considered. As we go towards the barrier from the higher energy side the experimental value of ratio increases from $\sim$1.5 to $\sim$2.5 in the interval $E_{c.m.}$/$V_{b}$ $\sim$0.94-0.82. We also compared the results with 1D BPM estimation and Wong's phenomenological prediction \cite{Wong}. The 1D BPM estimation in the no coupling mode was done using the code CCFULL; potential parameters for the $^6$Li+$^{28}$Si system  were taken from our previous work \cite{man07}. The 1D BPM analysis conforms to the experimental data at higher energies but underestimates at sub barrier region. In the case of $^6$Li+$^{28}$Si system, the values of the parameters used in the Wong model are $V_b$= 6.87 MeV and $R_b$= 8.02 fm \cite{Vaz} and curvature ($\hbar$$w$)= 3.24 is taken from Wong parameterization \cite{Wong}. The ratio from Wong model showed decreasing tendency towards the barrier.

We have included the ratio data of Ref.~\cite{Pakou09} in Fig.~\ref{7Li+6Li-coup} for comparison. It is obvious that the trend of {$\sigma_{fus}$}({$^6$Li})/ {$\sigma_{fus}$}({$^7$Li}) does not match the trend of the ratio obtained with the present measurement. The former shows an increasing ratio as the energy decreases below the barrier. However it is to be emphasized that the fusion cross section for $^6$Li+$^{28}$Si at low energies estimated from the $^{32}$S residue channel only, corroborates quite nicely with the fusion cross section estimates for $^6$Li+$^{28}$Si of Ref.~\cite{Pakou09}. In Ref.~\cite{Pakou09}, compared to present measurement method, the fusion cross section has been derived using a different technique. The observed mismatch in the data for {$\sigma_{fus}$}({$^6$Li})/{$\sigma_{fus}$}({$^7$Li}), plotted in Fig.~\ref{7Li+6Li-coup}, points to the underestimation of the derived fusion data of $^7$Li+$^{28}$Si in Ref. \cite{Pakou09}. Similar ratio was obtained for the systems $^{6,7}$Li+$^{59}$Co \cite{Cbeck,Adiaz5} and $^{6,7}$Li+$^{24}$Mg \cite{Mray} and are plotted in the same Fig.~\ref{7Li+6Li-coup}. To illustrate the contents of this figure more quantitatively we present, in Table \ref{table2}, the ratio values with errors and available barrier parameters ($V_b$, $R_b$, $\hbar\omega$) for each of the reactions. It is seen from the table that the nature of variation of the ratio is similar for $^{28}$Si (our data) and $^{24}$Mg. In the case of $^{59}$Co, though the behaviour is same, the slope of the ratio appears to be steeper in the neighbourhood of barrier compared to earlier cases. But for the results of ref.~\cite{Pakou09} ratio increases more rapidly below the barrier.

\begin{figure}
\resizebox{0.5\textwidth}{!}
{\includegraphics{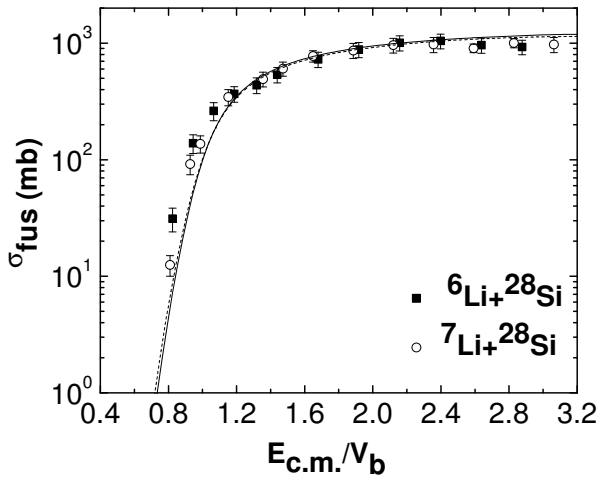}}
\caption{Measured $\sigma_{fus}$ plotted against $E_{c.m.}$/$V_{b}$ for the two systems, $^6$Li+$^{28}$Si (solid squares) and $^7$Li+$^{28}$Si (open circles). The solid and dashed curves represent the 1D BPM calculations for the $^7$Li and $^6$Li projectile systems respectively.}
\label{6,7Li-new1DBPM}      
\end{figure}

\begin{figure}
\resizebox{0.5\textwidth}{!}
{\includegraphics{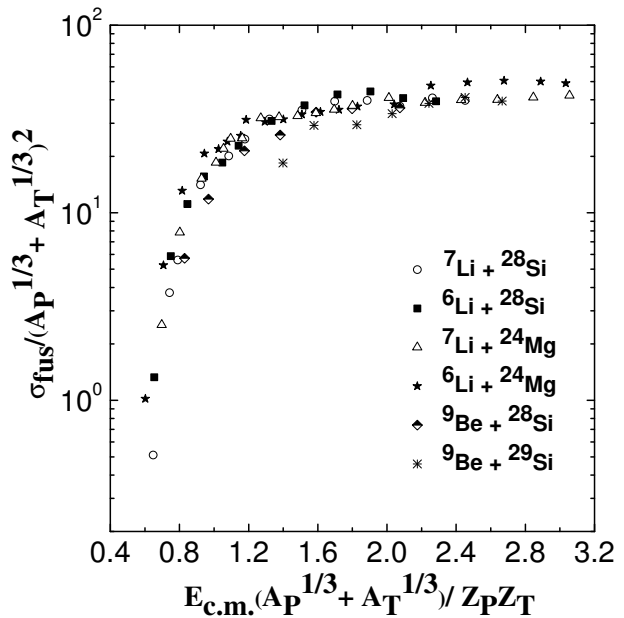}}
\caption{ Reduced fusion cross section of the system $^6$Li+$^{28}$Si alongwith other neighbouring target-projecile combinations.}
\label{Fus-red}      
\end{figure}

In Fig.~\ref{Fus-red} we have shown the comparison of our present fusion data for $^6$Li+$^{28}$Si with those of nearby target- projectile systems viz., $^7$Li+$^{28}$Si \cite{man07,man08}, $^{6,7}$Li+$^{24}$Mg \cite{Mray}, $^{9}$Be+$^{28}$Si \cite{Hugi}, $^{9}$Be+$^{29}$Si \cite{fig}, in the reduced scale, following the prescription of Gomes $et$ $al.$ \cite{Gomes}. In the higher energy region all the above systems showed almost similar type of fusion excitation. However at below the barrier the fusion cross sections are not identical for all the systems. We observed that fusion with more loosely bound nuclei $^6$Li is more compared to $^7$Li at below barrier energies for both the targets $^{28}$Si and $^{24}$Mg presumably due to lower breakup threshold of $^6$Li compared to that of $^7$Li.

\begin{table*}
\caption{\label{table2} Ratios of measured fusion cross sections for $^6$Li and $^7$Li projectiles with targets i.e., $^{28}$Si (present), $^{59}$Co \cite{Cbeck,Adiaz5}, $^{24}$Mg \cite{Mray} and $^{28}$Si \cite{Pakou09} as a function of $E_{c.m.}$/$V_{b}$ together with barrier parameters.}
\vglue 0.4cm
\begin{center}
\begin{tabular}{|c|c|c|c|c|c|c|c|}
\hline 
& & & & & & &\\
System&$E_{c.m.}$/$V_{b}$&{$\sigma_{fus}$}({$^6$Li})/ {$\sigma_{fus}$}({$^7$Li}) &Error ($\pm$)&System &$V_b$(MeV)& $R_b$(fm)&$\hbar\omega$ \\
\hline
& & & & & & &\\
$^{6,7}$Li+$^{28}$Si&0.82 &2.49 & 0.75 & & & &\\
(present)&0.94 &1.50 &0.39  &$^{6}$Li+$^{28}$Si&6.87 &8.11 &3.20\\
& 1.17 &1.06 &0.23  &$^{7}$Li+$^{28}$Si& 6.79& 8.22 &2.97\\
&1.33 & 0.89& 0.20& & & &\\
&1.45 & 0.88& 0.18& & & &\\
&1.66 & 0.94& 0.17& & & &\\
&1.90 & 1.02& 0.21& & & &\\
&2.14 & 1.05& 0.22& & & &\\
&2.37 & 1.07& 0.22& & & &\\
&2.85 & 0.92& 0.16& & & &\\
& & & & & & &\\
$^{6,7}$Li+$^{59}$Co &0.90 &2.35 & 0.51 & & & &\\
(Ref. [10,23])&0.93  &1.58 &0.53 &$^{6}$Li+$^{59}$Co&12.0 &7.6 &8.1\\
& 1.08 &1.99 &0.66  &$^{7}$Li+$^{59}$Co& 11.3& 7.5 &4.2\\
&1.15 & 1.06& 0.38& & & &\\
&1.23 & 1.08& 0.38& & & &\\
&1.31 & 0.73& 0.26& & & &\\
&1.38 & 0.85& 0.29& & & &\\
&1.46 & 0.88& 0.30& & & &\\
&1.54 & 1.06& 0.36& & & &\\
&1.84 & 1.10& 0.39& & & &\\
&1.92 & 0.84& 0.28& & & &\\
&1.99 & 0.95& 0.32& & & &\\
& & & & & & &\\
$^{6,7}$Li+$^{24}$Mg &0.83 &1.99 & 0.52 & & & &\\
(Ref. [25])&0.95  &1.59&0.37 &$^{6}$Li+$^{24}$Mg& 6.48&- &-\\
& 1.10 &1.31 &0.30  &$^{7}$Li+$^{24}$Mg& 6.36&-  &-\\
&1.20 & 1.13& 0.26& & & &\\
&1.28 & 0.92& 0.21& & & &\\
&1.37 & 0.98& 0.22& & & &\\
&1.64 & 0.93& 0.19& & & &\\
&1.76 & 0.97& 0.20& & & &\\
&1.89 & 0.97& 0.20& & & &\\
&2.02 & 0.96& 0.20& & & &\\
&2.14 & 0.95& 0.20& & & &\\
&2.39 & 0.88& 0.18& & & &\\
&3.38 & 1.16& 0.24& & & &\\
&3.59 & 1.11& 0.23& & & &\\
& & & & & & &\\
$^{6,7}$Li+$^{28}$Si&0.84 &2.79 & 0.34 & & & &\\
(Ref. [26])&0.87  &3.02 &0.32 &$^{6}$Li+$^{28}$Si&7.03 & -&-\\
& 0.97 &2.13 &0.32  &$^{7}$Li+$^{28}$Si& 6.95&- &-\\
&1.17 & 1.41& 0.22& & & &\\
&1.48& 1.20& 0.24& & & &\\
&1.58 & 1.19& 0.24& & & &\\
\hline
\end{tabular}
\end{center}
\end{table*}

\section{Summary and conclusion}

We have measured the total fusion cross section for $^6$Li+$^{28}$Si, at sub and above barrier energies viz., at $E_{lab}$= 7-24 MeV by characteristic $\gamma$-ray method. The total fusion cross sections at above barrier energies were extracted from the measured total $\gamma$-ray cross section. The measured $\sigma_{fus}$ showed good agreement with 1D BPM up to 2$V_b$ but small suppression was found beyond the 2$V_b$ region similar to our earlier observation in the case of $^7$Li+$^{28}$Si system. In the sub barrier region, the $\sigma_{fus}$ was extracted from total $\gamma$-ray cross section and also from summing the channel cross section. The results using the two different approaches are consistent and show enhancement with respect to 1D BPM prediction at below the Coulomb barrier region. The coupled channel calculation considering rotational coupling of 2$^+$ (1.779 MeV) state of target $^{28}$Si does increase the fusion yield by a small amount in the sub-barrier region but does not describe the observed enhancement completely. The enhancement is found to be larger for $^6$Li compared to $^7$Li at these energies. The ratio {$\sigma_{fus}$}({$^6$Li})/{$\sigma_{fus}$}({$^7$Li}) predicted from Wong and 1D BPM estimations falls short of experimental values in the same range. 

In our analysis we could not take into account the effects of coupling to breakup/transfer channels as we could not identify these components, nor could we separate the ICF events. The possible factors causing the increased sub-barrier yield may be due to these effects. Though $n$-transfer possibilities have been explored by Pakou $et$ $al$.
\cite{Pak1}, the following studies \cite{Cbk7,mdp,Cbk8} showed that in spite of small breakup cross section ICF arising out of coupling to BU/TR may be appreciable. Very recently F.A. Souza $et$ $al$. \cite{Souza,Souza1} have experimentally demonstrated the dominant occurrence of $d$-ICF for $^6$Li+$^{59}$Co. It is interesting to note  from  Fig.~\ref{7Li+6Li-coup}, that the behaviour of {$\sigma_{fus}$}({$^6$Li})/{$\sigma_{fus}$}({$^7$Li}) for $^6$Li+$^{24}$Mg, $^{28}$Si, $^{59}$Co (except the data point close to $E_{cm}$/$V_b$ $\sim$1) is similar, indicating the dominance of identical reaction mechanisms around this energy region. The observed systematic behaviour of the presented data in Fig.~\ref{7Li+6Li-coup} prompted us to believe that for $^6$Li+$^{28}$Si also, the $d$-ICF/$d$-transfer may be the dominant process. However it should be noted that the characteristic $\gamma$-ray detection technique for fusion measurement can not really identify the reaction mechanism producing a particular residue. To identify the mechanism, exclusive coincidence measurement is absolutely necessary. In this context, we feel that the particle-gamma coincidence like $\alpha$-$\gamma$ coincidence measurement is not sufficient enough to distinguish the $d$-ICF and single nucleon transfer processes. Both these processes will have $\alpha$-particle in the exit channel. For $d$-ICF, $\alpha$ in the exit channel will come from breakup of $^6$Li and for nucleon transfer process it will come from breakup of either $^5$Li or $^5$He. Exclusive particle-particle coincidence, for instance exclusive coincidence between $\alpha$ and proton coming from $^5$Li breakup, can tag the single nucleon transfer process.

\section{Acknowledgement}

 The authors would like to thank the members of the TIFR/ BARC (Mumbai) and of IOP (Bhubaneswar) pelletron facilities for their sincere help and cooperation throughout the experiments. We also would like to thank P.K. Das, K. Chatterjee, B.P. Das, J. Panja and S. Chatterjee of Saha Institute of Nuclear Physics for their technical support during the experiments and during preparation of target.

\end{document}